\documentclass[
    ,final            
  ]
  {aipproc}

\layoutstyle{6x9}
\usepackage{graphicx}

\begin{document}

\title{Electromagnetic Signals at RHIC}



\author{Simon Turbide and Charles Gale}{
  address={Department of Physics, McGill University, 
3600 University Street, Montreal, Canada H3A 2T8}
}

\begin{abstract}
We calculate the direct photon yield in central and mid-peripheral Au+Au collisions at the Relativistic Heavy-Ion Collider (RHIC) ($\sqrt{S_{NN}}=200$GeV).  The processes involving the 
propagation of jets have been convolved with a leading order treatment of jet energy loss 
in the medium and a one dimensional hydrodynamic expansion.  The quark-gluon plasma (QGP) 
contribution turns out to be important, especially the in-medium conversion of a 
jet into a photon, for successfully describing recent photon measurements.
\end{abstract}

\maketitle


The search for the creation of a quark-gluon plasma (QGP) in relativistic heavy ion collisions has driven many experimental and theoretical investigations.  Electromagnetic radiation 
(real and virtual photons), as it does not suffer final state interactions, constitutes
a class of very interesting probes of the hot and dense medium created during those 
collisions.  Indeed, they carry to the detectors information about the state of the system at 
the time they have been produced.  If a hot thermalized phase is created in 
nucleus-nucleus collisions, we thus expect to see in the photon spectrum, after subtracting 
the background coming from the decay of neutral mesons, some trace of these thermal photons.

The first direct photon measurements for Pb(158$A$ GeV)+Pb at the Super Proton synchrotron (SPS)
have been done by the WA98 collaboration~\cite{SPS}.  Our previous calculations~\cite{Turb:04} 
assert that with a moderate value of the initial temperature ($T_i=205$ MeV), there 
was no strong evidence of the formation of the QGP. In other words, a contribution from the QGP was 
not crucial to a successful interpretation of the data.  At RHIC energies, however, the discovery of 
high $p_T$ suppression in hadron spectra~\cite{phenix2} is highly suggestive of the formation 
of a hot and dense phase. This represents new physics, over that of the SPS.  The 
most currently plausible explanation for this observation is 
the energy loss of jets by induced gluon bremsstrahlung, as they go through a hot and dense 
partonic phase.  In analogy with the jets interacting strongly with the partonic phase, 
it has been suggested in ~\cite{prlphoton} that jets could interact electromagnetically 
with the medium as well.  It was found that the direct annihilation and Compton scattering of 
quark jets, with incoming thermal partons, was an important source of photons at 
intermediate $p_T$.  However, jet energy loss still needed to be included consistently. 

Before  a comparison with experimental data, we briefly present the different 
contributions to the direct photon spectrum.   Firstly, direct photons from primary 
hard Compton and annihilation processes, at the partonic level, $a+b\to c+\gamma$, are 
produced during the overlap of the nuclei, with no final state effect.  Secondly, jets 
also produced during the initial early stages ($a+b\to c+d$), can thereafter go 
through the medium, lose some energy by gluon bremsstrahlung ($c\to c^*+g$), and 
fragment into photons outside the medium ($c^*\to c^*+\gamma$). This jet-fragmentation 
scenario can be expressed by
\begin{equation}
\frac{dN^{jet-frag}}{d^2p_T dy}=\sum_f \frac{dN^f}{d^2p_Tdy}\otimes D_{f/\gamma}(z,p_T)
\end{equation}
where $D_{f/\gamma}$ is the photon fragmentation function, and $dN^f/d^2p_Tdy$ is the 
momentum distribution of the parton of flavour $f$, after it has crossed the hot medium.  
The path length dependence of this distribution is obtained with the complete leading order 
description of Arnold, Moore and Yaffe (AMY)~\cite{AMY}. The primary hard direct photons 
added to the photons from fragmentation, are defined to be the prompt photons. 

As previously mentioned, the interaction of jets with the medium can also produce photons 
by direct scattering (jet+$i\to j+\gamma$), where $(i,j)$ stand for thermal partons, but 
also through  medium-induced photon bremsstrahlung.  The photon yield generated by those 
jet-induced mechanisms can be written by
\begin{equation}
\frac{dN^{jet-induced}}{d^2p_T dy}=\int dt\int d^3x\int d^3q \frac{dN^q}{d^2q_Tdy^{'}}(t)\otimes \sigma^{jet-induced}_{jet(q)\to \gamma(p)}
\end{equation}
The cross-section for the direct process is dominated by transfer of the entire jet 
momentum to the photon, such that $\sigma^{jet-th}_{jet(q)\to \gamma(p)} \propto 
\delta^3(q-p)$: this process is labeled the jet-photon conversion.  Details about 
these mechanisms are given in Ref.~\cite{TGJM:05}.
Finally, the medium can also simply radiate photons.  Calculations for the 
QGP and hadronic gas induced radiations can be found in Refs.~\cite{guy3} 
and ~\cite{Turb:04} respectively.

The work here differs from that of Ref.~\cite{TGJM:05}, as we have extended our direct photon 
calculations to non-central collisions.  This is achieved by the following three 
steps.  (1) The initial jet spectrum is scaled with the overlap factor of the 
nuclei: $dN^f(b)= dN^f(b=0)T_{AB}(b)/T_{AB}(b=0)$.  (2) The thermalized medium 
evolves according to a 1-D Bjorken~\cite{Bjorken} expansion with the initial conditions 
given by $\tau_i T^3_i\propto dN(b)/dy/A_\perp(b)$, where $dN/dy$ is the particle rapidity 
density, and $A_\perp$ is the overlap area of the nuclei in the transverse plane. (3) 
Finally, the overlap zone of the nuclei has an ``almond`` shape, such that the 
total energy 
lose by a jet will depend on its direction.

\begin{figure}
  \includegraphics*[width=4in]{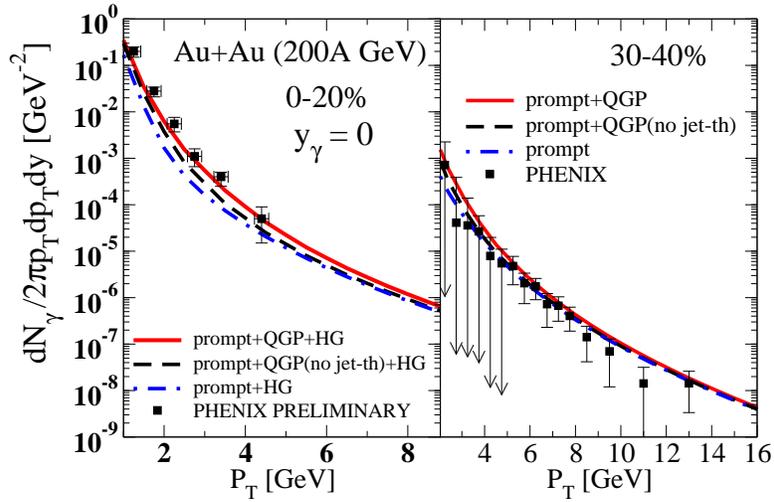}
  \caption{\label{fig:1} Direct photon spectrum at midrapidity, in Au+Au collision at RHIC. Left hand side: solid line, all direct photons; dashed line, all direct photon without the jet-photon conversion; and dot-dashed line, all direct photons without a QGP contribution; compared to PHENIX data~\cite{new_photon} for a 0-20$\%$ centrality class. Right hand side: solid line, prompt and QGP photons; dashed line, prompt and QGP without the jet-photon conversion; and dot-dashed line, prompt photons only; compared to PHENIX data~\cite{phe_pho} for a 30-40$\%$ centrality class.}
\end{figure}

Our photon yield calculations for Au+Au collisions at RHIC are shown in Fig.~\ref{fig:1}.  
For a fixed initial time $\tau_i=$0.26 fm/c, we 
use the initial temperatures $T_i=$370~\cite{Turb:04} and 340 MeV for central (0-20$\%$) and 
mid-peripheral (30-40$\%$) collisions respectively.  We define by QGP all photons which 
have been produced during the QGP phase.  For central collisions, we reproduce the preliminary 
data from PHENIX~\cite{new_photon} only when the QGP contributions 
are included, with our initial temperature, while  we are below the data by a factor $\sim$2 around $p_T=3$ GeV, when all QGP contributions, except the 
jet-photon conversion, are included. For the 30-40$\%$ centrality class (right panel), the PHENIX 
data~\cite{phe_pho} show good precision only at high $p_T$, such that we have not 
included the hadronic contribution, which is not expected to contribute much in this range.  
While the high $p_T$ spectrum seems to be fully dominated by prompt photons, our 
calculations show again an important contribution from the QGP in the intermediate 
$p_T$ window.  In proportion, this contribution is however $25\%$ less important 
than for the central class, as we might have expected, since the initial temperature and 
the size of the interacting zone both decrease with increasing impact parameter.

In conclusion, we argue that the QGP contributions are important, particularly 
the jet-photon conversion, to the interpretation of the
experimental data on direct photons.  We expect virtual photons to exhibit a 
similar feature: this topic is presently under study.

\begin{theacknowledgments}
 We thank S. Jeon and G.D. Moore for helpful discussions. This work was supported in part by the Natural Sciences and Engineering Research Council of Canada, and in part by le Fonds Nature et Technologies du Qu\'ebec.
\end{theacknowledgments}


\begin{thebibliography}{9}


\bibitem{SPS}
M. M. Aggarwal {\it et al.}, WA98 Collaboration, Phys. Rev. Lett. {\bf 85}, 3595 (2000).

\bibitem{Turb:04} 
S.~Turbide, R.~Rapp, and C.~Gale, 
Phys.\ Rev.\ C {\bf 69}, 014903 (2004).

\bibitem{phenix2}
S.S. Adler {\it et al.}, Phys. Rev. Lett. {\bf 91}, 072301 (2003).


\bibitem{prlphoton}
R.J. Fries, B. M\"uller and D.K. Srivastava, Phys. Rev. Lett {\bf 90}, 132301
(2003).

\bibitem{AMY}
P.~Arnold, G.~D.~Moore and L.~Yaffe, JHEP {\bf 0111}, 057 (2001); JHEP {\bf
0206}, 030 (2002).

\bibitem{TGJM:05}
S.~Turbide, C.~Gale, S.~Jeon and G.~D.~Moore,
Phys. Rev. C {\bf 72}, 014906 (2005).

\bibitem{guy3}
P. Arnold, G.D. Moore and L.G. Yaffe, J. High-Energy Phys. {\bf 0012}, 009
(2001).

\bibitem{Bjorken}
J.D. Bjorken, Phys. Rev. D {\bf 27}, 140 (1983).



\bibitem{new_photon}
S. Bathe, talk given at Quark Matter 2005, Budapest, Hungary (2005).

\bibitem{phe_pho}
S.S. Adler {\it et al.}, Phys. Rev. Lett. {\bf 94}, 232301 (2005).

\end{thebibliography}
\end{document}